\documentclass[superscriptaddress,twocolumn,amsmath,amssymb,aps,pra]{revtex4-1}

\usepackage{graphicx}
\usepackage{dcolumn}
\usepackage{bm}
\usepackage{xcolor}
\usepackage{hyperref}
\usepackage{graphicx}
\usepackage{amsmath}
\usepackage{bbold}
\usepackage[T1]{fontenc}

\newcommand{\beq}{\begin{equation}}
\newcommand{\eeq}{\end{equation}}
\newcommand{\beqs}{\begin{equation*}}
\newcommand{\eeqs}{\end{equation*}}
\newcommand{\beqa}{\begin{eqnarray}}
\newcommand{\eeqa}{\end{eqnarray}}
\newcommand{\beqas}{\begin{eqnarray*}}
\newcommand{\eeqas}{\end{eqnarray*}}
\def\bals#1\eals{\begin{align*}#1\end{align*}}
\def\bal#1\eal{\begin{align}#1\end{align}}

\newcommand{\bcent}{\begin{center}}
\newcommand{\ecent}{\end{center}}

\newcommand{\bitem}{\begin{itemize}}
\newcommand{\eitem}{\end{itemize}}

\usepackage[normalem]{ulem}

\begin{document}

\title{$Z_1$ oscillations and charge state in electronic 
stopping power from first principles} 

\author{Noor Ul Ain} 
\affiliation{Theory of Condensed Matter, Cavendish Laboratory, 
University of Cambridge, J J Thomson Avenue, Cambridge CB3 0US, 
United Kingdom}

\author{Emilio Artacho}
\affiliation{Theory of Condensed Matter, Cavendish Laboratory, 
University of Cambridge, J J Thomson Avenue, Cambridge CB3 0US, 
United Kingdom}
\affiliation{CIC Nanogune and DIPC, Tolosa Hiribidea 76, 20018
San Sebasti\'an, Spain}
\affiliation{Ikerbasque, Basque Foundation for Science, 
48011 Bilbao, Spain}

\date{\today}

\begin{abstract}
  The energy transfer rate from a projectile nucleus to the 
electrons of the matter it traverses depends on the charge of
that projectile, $Q=e Z_1$.
 At low projectile velocities the friction coefficient is 
known to oscillate with atomic number $Z_1$, since core 
electrons travel with the projectile screening its charge.
  The effective charge increases with velocity and oscillations 
disappear.
  That effect is studied here calculating electronic stopping
power from first principles for O and Mg projectiles shooting
through bulk Al, using real-time time-dependent density-functional
theory.
  Both projectiles represent maximum and minimum of the first
$Z_1$ oscillation, respectively.
  The oscillation is found to be very sensitive to the direction
of propagation, in spite of Al being quite an ideal metal for 
many purposes.
  The critical velocity for the oscillation disappearance ranges
between below 0.1 a.u. and beyond 1 a.u. for the explored 
trajectories.
  The charge state is independently quantified with Hirshfeld and
Voronoi analyses, offering remarkably consistent results 
in spite of their very different partition methods, as well as
with an effective definition based on the stopping power itself.
  They display a gradual undressing of the projectile's core 
electrons with increasing velocity in qualitative accordance 
with expectations.
  However, electron density plots in real space present a richer 
picture in which the undressing is partly due to the electrons
trailing behind the projectile, suggesting possible 
phenomenological descriptions correcting for the deformation
of the density in terms of multipoles beyond 
the net charge.
  The plots also offer insights into dissipation by core electrons.
\end{abstract}

\maketitle


\section{Introduction}

  The interaction of charged projectiles with matter has been a 
significant area of interest since the discovery of subatomic particles
\cite{Bragg1905,Rutherford}.
  It is behind various important processes in condensed matter science,
such as collisions, ionization, scattering, sputtering, and radiation 
damage, of applied relevance to various fields as nuclear power 
generation, aerospace engineering, ion implantation, nuclear medicine 
and ion radiotherapy. 
  The kinetic energy of the projectile dissipates as it passes through matter.
  The amount of energy lost per unit distance is referred to as 
stopping power, $S = - \frac{dE}{dx}$.
  Energy loss occurs because of the interaction of the projectile with 
the host nuclei and electrons, the stopping power being expressed as
the sum of the nuclear $S_n$ and electronic $S_e$ components \cite{Sigmund}. 
  Electronic stopping dominates at high velocities, typically 
above $\sim 0.1$ a.u.

\begin{figure}[t!] 
\includegraphics[width=0.35\textwidth]{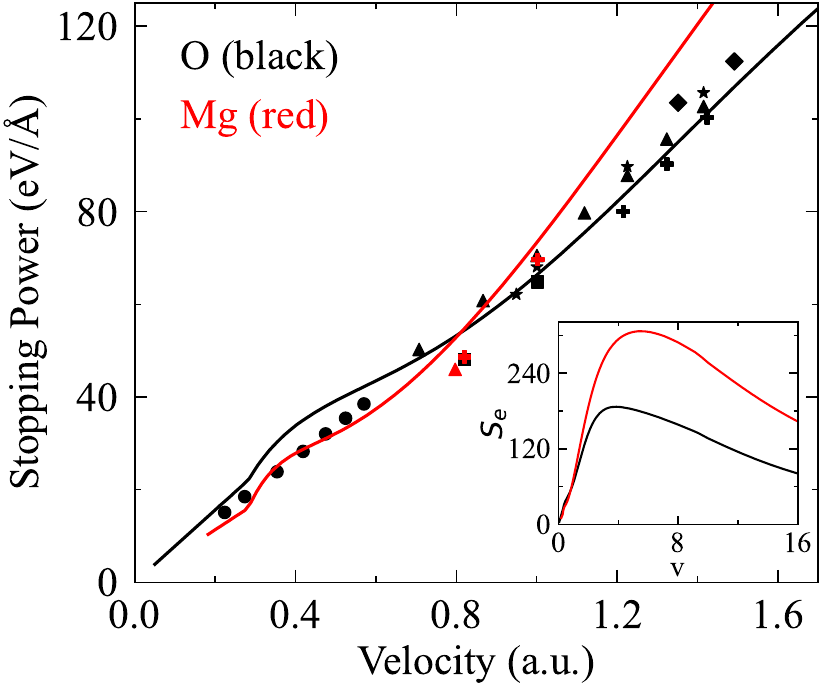}
\caption{Electronic stopping power $S_e$ vs projectile velocity from 
the literature. 
Black: $S_e$ for O, SRIM 2013~\cite{SRIM} (continuous line) and experiments
(circles~\cite{Ormrod1965}, triangles~\cite{Santry1992}, stars~\cite{Porat1961},
squares~\cite{Ward1979}, crosses~\cite{Xiting2000}, and rombi~\cite{Abdesselam1992}).
Red: $S_e$ for Mg, SRIM 2013 and experiments (crosses~\cite{Ward1979} and 
triangle~\cite{Lennard1987}).}
\label{fig:SRIM}
\end{figure}

  Although linear response predicts the electronic stopping power scaling
as $Q^2$, being $Q$ the charge of the projectile~\cite{lindhard1954}, 
the low-velocity $S_e$ actually displays an oscillatory behaviour 
as a function of the atomic number of the projectile ($Z_1$).
  This phenomenon, first predicted from theory \cite{Ormrod1965} 
and then observed in experiments \cite{Eisen1968,Puska1983,Sortica2019}, 
is due to the fact that core electrons move with the projectile 
thereby screening its nucleus charge, the oscillation essentially 
following the periodic table \cite{Echenique1990,Briggs_1973}.

  When increasing the velocity, however, deeper electrons start to
get involved in the dissipation and eventually higher $Z_1$ leads 
to higher $S_e$, giving a crossing of those $S_e(v)$ curves that
started off with the reverse.
  Fig.~\ref{fig:SRIM} illustrates it in the case of O ($Z_1=8$)
and Mg ($Z_1=12$) projectiles in bulk Al, prototypical cases of 
maximum and minimum of the $Z_1$ oscillation in a simple metal.
  The empirical $S_e(v)$ curves given by SRIM~\cite{SRIM} show 
how $S_e$ starts higher for O, and then the behaviour is
inverted with a crossing at $v\sim 0.8$ a.u.
  It is corroborated by the experimental values 
\cite{Porat1961,Ormrod1965,Ward1979,Abdesselam1992,Santry1992,
Xiting2000,Lennard1987}, also shown in the figure. 
  It can be seen in terms of the effective charge of the 
projectile defined as \cite{Echenique1990},
$Z_1^* \equiv \sqrt{S_e(Z_1) / S_e(\mathrm{H})}$,
referring to the stopping power of a proton projectile, H.
  $Z_1^*$ essentially reflects the projectile's valence charge 
at low velocity, starting as a fairly constant value and then 
gradually increasing with velocity when the core electrons 
become active in the stopping process.

  The aim of this work is furthering the understanding of the relation
between stopping and the actual charge around the projectile.
   We use the explicit first-principles electronic evolution
technique started in 2007 \cite{Pruneda2007,Krasheninnikov2007,
mason2007}, which has been then successfully applied to quite
varied systems (see e.g. the review in \cite{Correa2018}).
  In addition to the first-principles calculations of $S_e$,
it allows analysing the dynamical electron density and the 
dependence of both with projectile's velocity and trajectory.
  After describing the method, results are presented for 
$S_e(v)$, $Z_1^*$, charge distribution, and various measures 
of the dynamical charge for O and Mg in bulk Al, and their 
dependence with impact parameter.

 
\begin{figure}[t!] 
\includegraphics[width=0.35\textwidth]{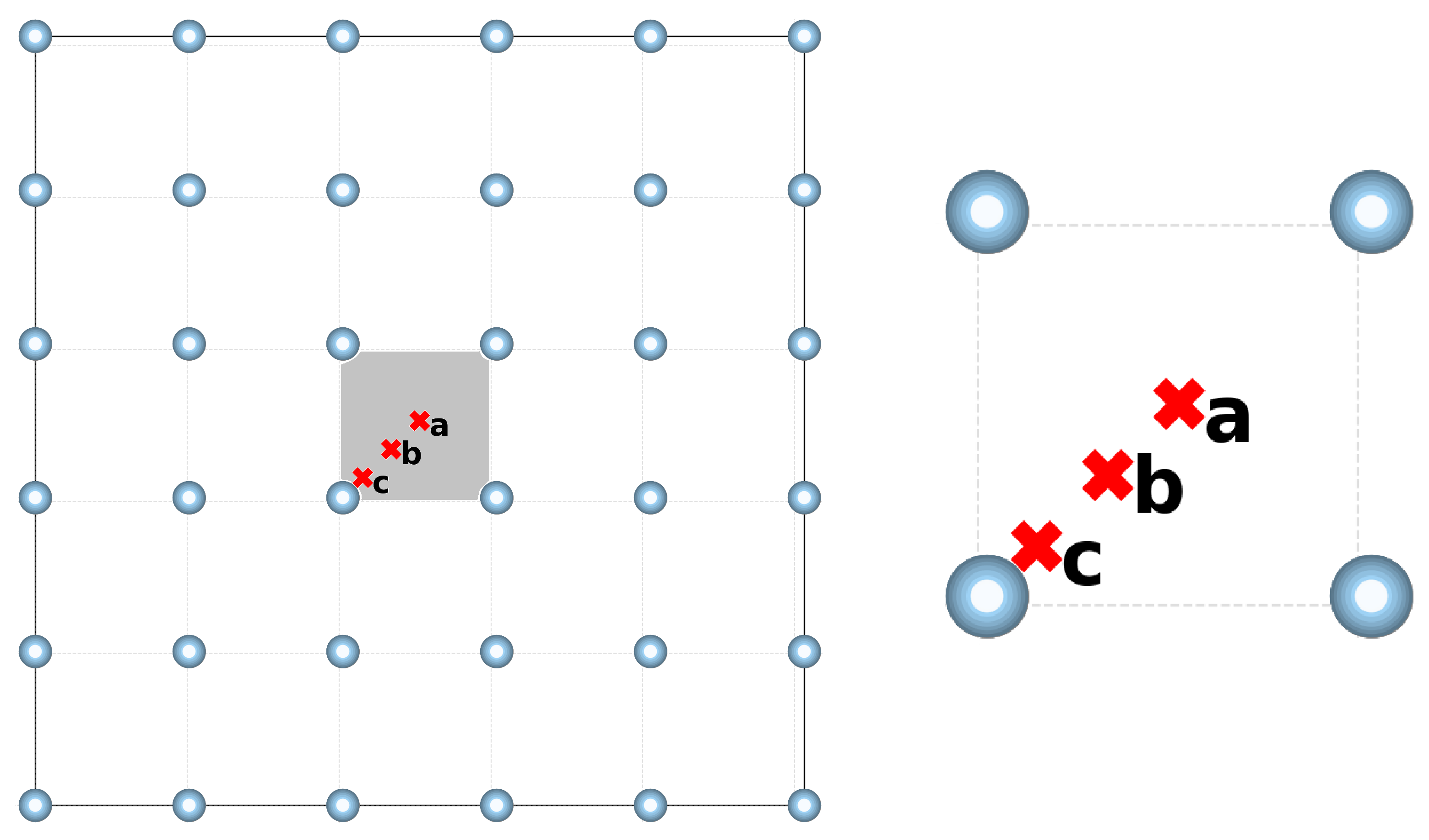} \\
\includegraphics[width=0.42\textwidth]{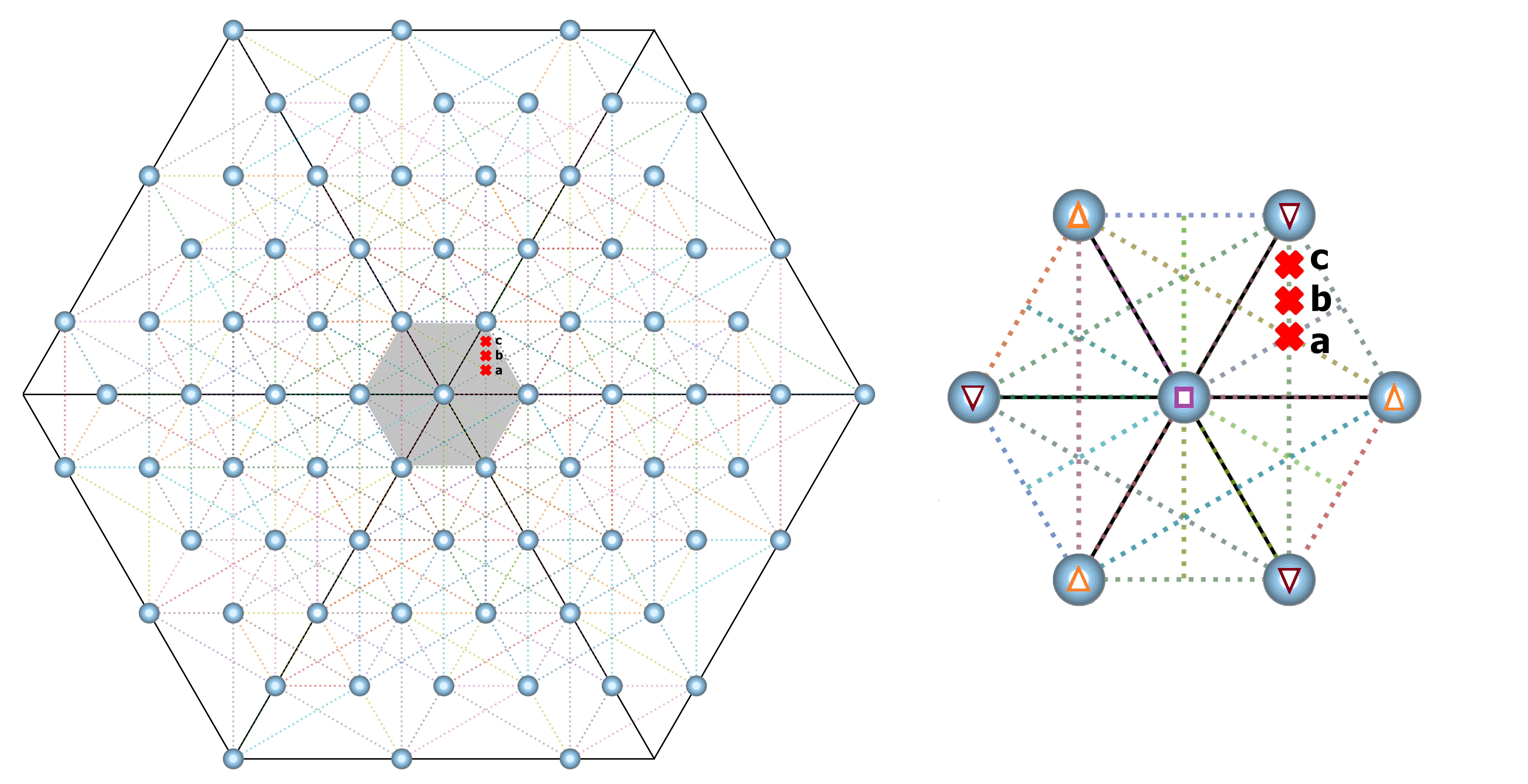}
\caption{
  Normal view down the $\langle 001\rangle$ and $\langle 111\rangle$ 
channel in the upper and lower panel, respectively.
  Cross signs represent the projectile’s initial position 
for the sampled trajectories for each channel.
  Note that Al nuclei defining the channel cross section
are not within the same plane, as illustrated in the 
lower panel with up-triangles, down-triangles and squares
being on different planes.}
\label{fig:trajectories}
\end{figure} 


\section{Method}
\label{sec:method}

  Real-time time-dependent density-functional theory (RT-TDDFT) 
\cite{Runge1984} was used to calculate the evolution of the electrons 
in a supercell of bulk Al under the passage of oxygen and magnesium 
projectiles following various constant-velocity trajectories. 
  The method was first described in \cite{Pruneda2007}.
  The initial Kohn-Sham (KS) single-particle states are obtained from a 
conventional, static DFT calculation of the supercell with the 
projectile located at its initial position.
  The KS-RT-TDDFT evolution is then obtained by discretising time, and 
moving the projectile as the electronic wave functions are propagated.
  The functional by Perdew, Burke and Ernzerhof (PBE) \cite{Perdew1996} 
is used as exchange-correlation (XC) potential assuming adiabatic XC
for the TD-DFT evolution.
  The {\sc Siesta} method \cite{Soler2002,Garcia2020} is used for the 
calculations, which is based on numerical atomic orbitals as basis sets.
  We used the TDDFT {\tt new\_integrator} branch of the {\sc Siesta}
4.1.5 release. 
  It is available at gitlab 
\footnote{https://gitlab.com/sanaz.grivani89/TDDFT.git}
pending merge to the main {\sc Siesta} development trunk
(it will be included in the {\sc Siesta} 5.6 release).

  The simulations were performed for a bulk aluminium $3\times 
3\times 3$ supercell of the conventional face-centered cubic (FCC) 
cell, using the 4.05 \AA\ experimental lattice constant.
 The box contains 108 atoms of Al plus the projectile.
 Real-space density and potential are expressed in a discretised 
grid characterised by a mesh cut-off of 200 Ry \cite{Soler2002}.
 Reciprocal-space discretisation is provided by a $2\times 2\times 2$ 
$\mathbf{k}$-point grid displaced to the centre of the 
discretisation cell, giving an effective $\mathbf{k}$-grid cut-off
of 12.15 \AA\ \cite{Moreno1992}.
  Core electrons were replaced by pseudopotentials of the optimised 
norm-conserving Vanderbilt (ONCV) kind \cite{Hamann2013}, 
considering all electrons explicitly as valence except for the 
$1s$ cores.
  The pseudopotentials for Al, Mg and O were obtained from 
the PseudoDojo repository~\cite{pseudodojo2018}.
  The single-particle Kohn-Sham (KS) wave functions are expanded 
in a basis of numerical atomic orbitals. 
  A double-$\zeta$ polarized basis is used for valence and semicore
shells.
  Radial wave functions for the first $\zeta$ were generated from 
the DFT solution for the pseudo atom under a smooth confining 
potential \cite{Junquera2001}, and for the following $\zeta$'s 
using the split-valence scheme described in \cite{Soler2002}.
  Further details on the basis including smooth 
confinement~\cite{Junquera2001} parameters
and various cut-off radii are provided in the Appendix. 

  With the projectile at its initial position, a conventional 
static DFT calculation is performed for the supercell. 
  A tolerance for the maximum deviation in density-matrix 
elements of $10^{-4}$ is used for self-consistency. 
  The occupied KS eigenvectors in this calculation become the 
initial KS wave functions to be propagated in TDDFT.
  For the latter, time is discretised in steps of 0.9 attoseconds 
(see convergence tests in the Appendix)
and the coefficients of the KS expansion are evolved using
the Crank-Nicholson integrator adapted to an evolving basis 
set: 
  Since the basis functions move with the atoms, the Hilbert 
subspace spanned by the basis rotates with atomic motion
\cite{Artacho2017}, giving a modified Hamiltonian for an 
asymptotically unitary integration \cite{Halliday2021}.

  Host Al nuclei are kept frozen, thereby obtaining the purely
electronic component of the stopping power. 
  Allowing them to move makes a very small difference, 
since the projectile crosses the sample box in a few femtoseconds
at most, a very short time for nuclear dynamics in the absence
of head-on collisions.
  The projectile is set on a constant velocity (vector) trajectory,
as commonly done in stopping power calculations \cite{Halliday2022}.
  It allows establishing $S_e$ for a well defined velocity
from the energy rise as the projectile is dragged through.

  Several trajectories are sampled in this work, namely,
two sets of channelling ones, along $\langle 001\rangle$ and
$\langle 111\rangle$ channels, and an averaging trajectory 
that follows the $(1,\phi,\phi^2)$ 
incommensurate direction along which the system is not periodic
\cite{Schleife2015,Quashie2016}, $\phi$ being the golden ratio 
$(1+\sqrt{5})/2$.
  The channelling ones are of different impact parameter 
(closest distance to the nuclei defining the channel) and are 
presented in Fig.~\ref{fig:trajectories}. 
  Each set contains one along the centre of the corresponding 
channel (hyperchannelling), called $a$, with impact parameters
of 1.432 \AA\ and 0.955 \AA\ for $\langle 001\rangle$ and 
$\langle 111\rangle$, respectively, and two more, $b$ and $c$, 
with impact parameters 0.902 \AA\ and 0.371 \AA\ for 
$\langle 001\rangle$, and 0.672 \AA\ and 0.389 \AA\ for 
$\langle 111\rangle$.

\begin{figure}[h!] 
\includegraphics[width=0.35\textwidth]{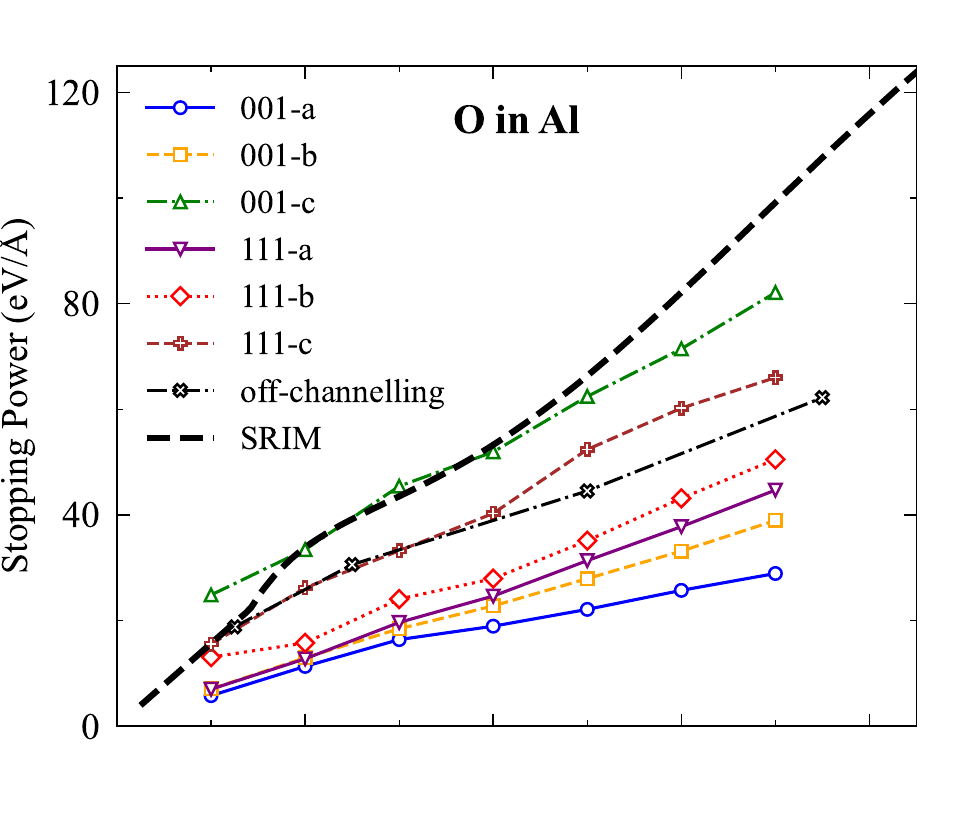}\\
\vspace{-15pt}
\includegraphics[width=0.35\textwidth, trim=0 0 0 8mm,clip]{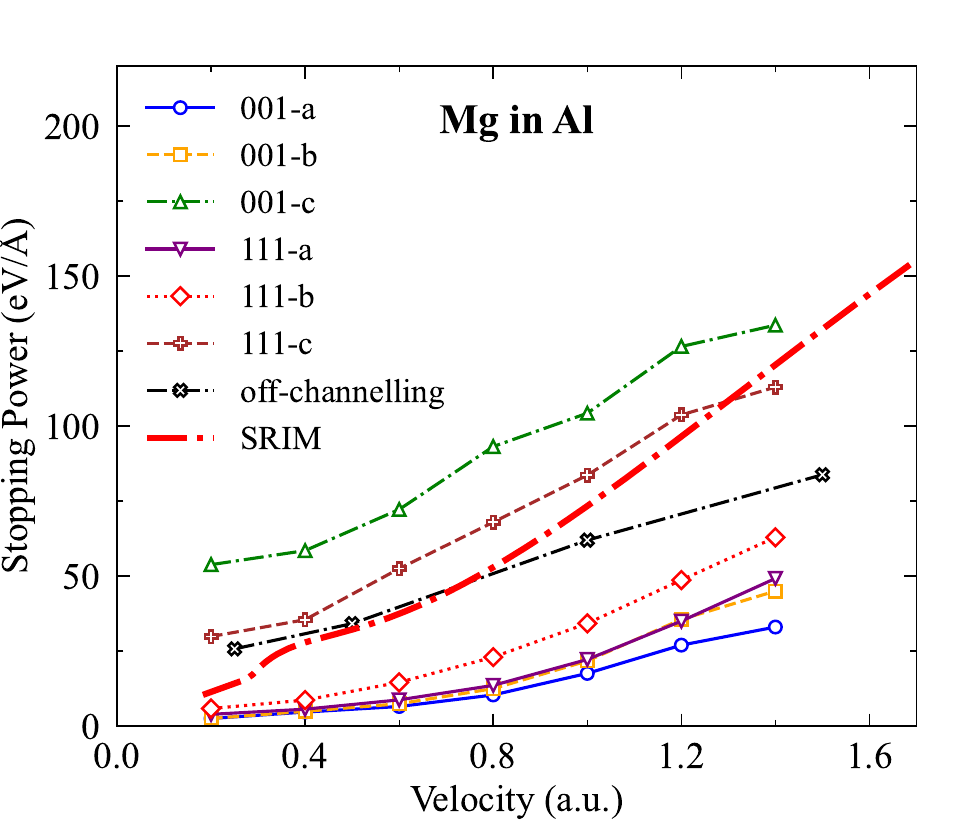}
\caption{Electronic stopping power $S_e$ for O (upper panel)
and Mg (lower panel) projectiles through bulk Al, including
SRIM results and this work's results along different 
trajectories as noted in the key referring to nomenclature
established in Fig.~\ref{fig:trajectories}.
  Off-channelling refers to the incommensurate trajectory
described in the text.}
\label{fig:stopping}
\end{figure} 


\section{Results and Discussion}
\label{sec:results}

\subsection{Electronic Stopping Power}
\label{sec:ESP}

\begin{figure}[h!] 
\includegraphics[width=0.35\textwidth]{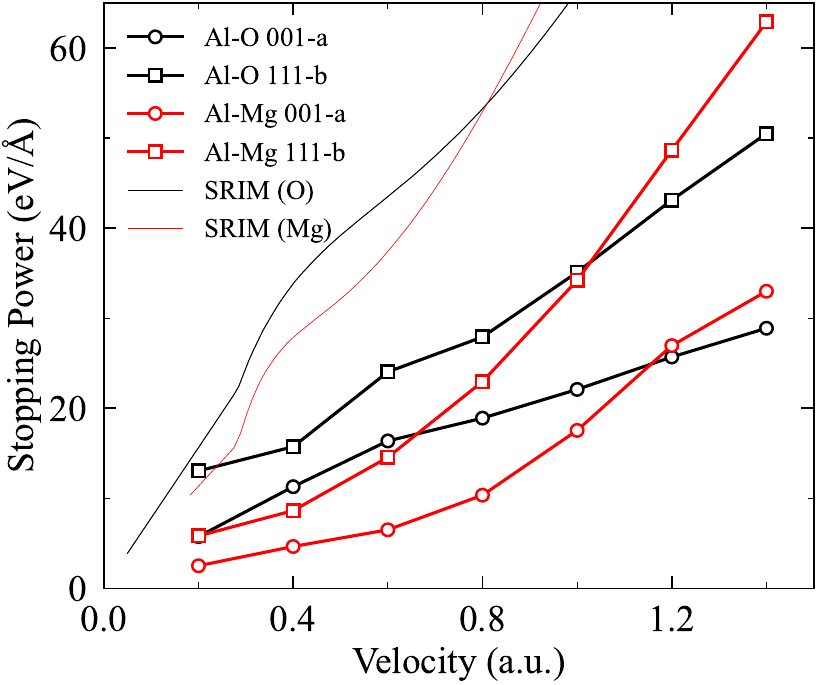}\\ 
\vspace{-3pt}
\caption{Crossover in electronic stopping power $S_e(v)$ 
curves for O (black) and Mg (red) projectiles.}
\label{fig:crossings}
\end{figure}

  Results for electronic stopping power versus velocity, $S_e(v)$,
are shown in Fig.~\ref{fig:stopping} for all trajectories of
oxygen (upper panel) and magnesium (lower) projectiles across
bulk aluminium.
  The corresponding SRIM curves are also shown for comparison.
  The first point to note is the significant trajectory dependence,
shown by the spread of the curves, giving a distribution width for
stopping power of a similar size as its average.
  Although not unexpected, it is not found in reference 
calculations when discussing $Z_1$ 
calculations~\cite{echenique86}, many of which are done 
for a homogeneous electron liquid (jellium) as host,
and for which all trajectories are equivalent by symmetry.
  Fig.~\ref{fig:crossings} shows the crossings of
the O and Mg $S_e(v)$ curves for two trajectories, namely, 
(001)-a and (111)-b.
  Although it is the expected behaviour the crossing point 
changes appreciably.

\begin{figure}[t!] 
\includegraphics[width=0.35\textwidth]{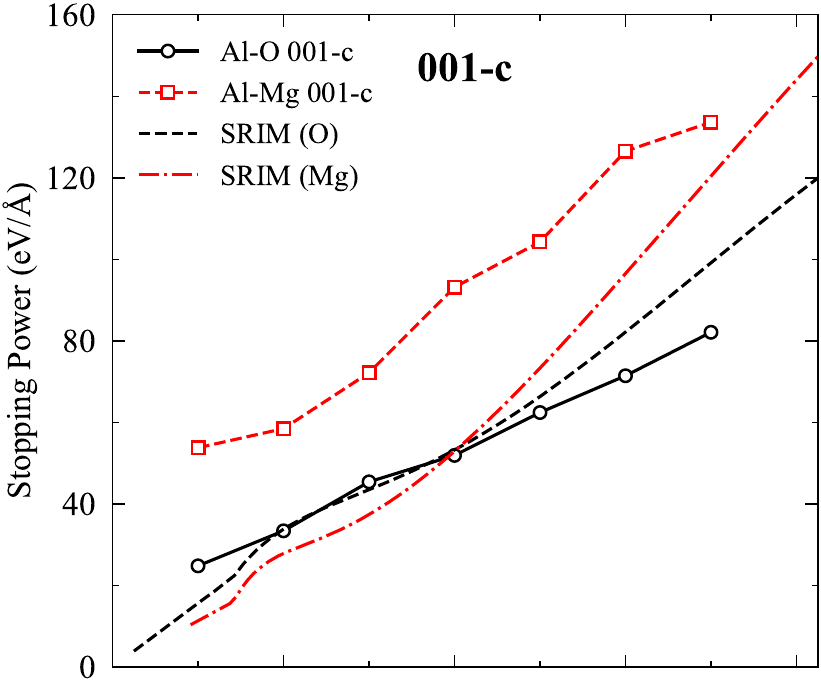} 
\vspace{-6pt}
\includegraphics[width=0.35\textwidth]{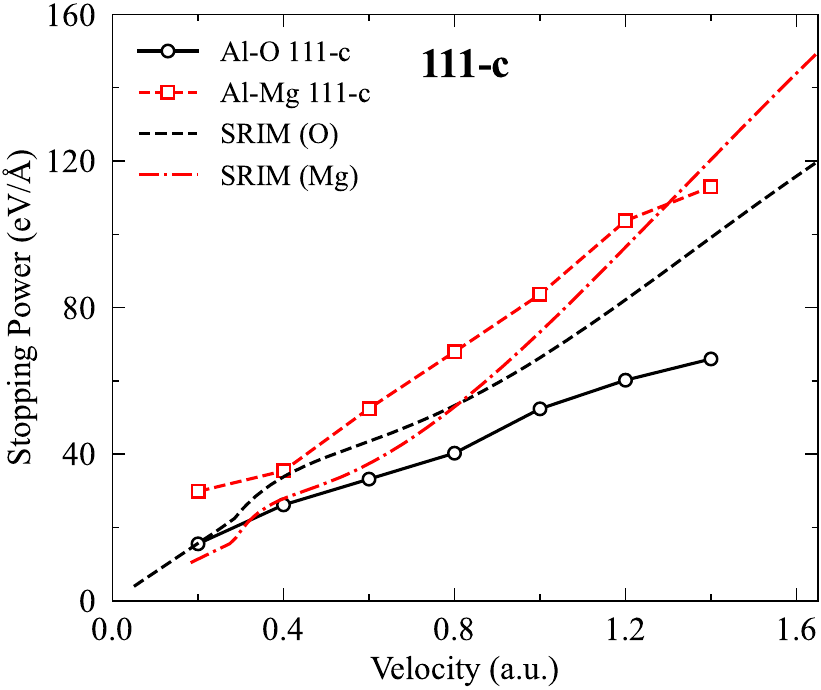}\\ 
\caption{Cases of $S_e(v)$ curves displaying no
crossover.}
\label{fig:non-crossings}
\end{figure}

\begin{figure}[h!] 
\includegraphics[width=0.35\textwidth]{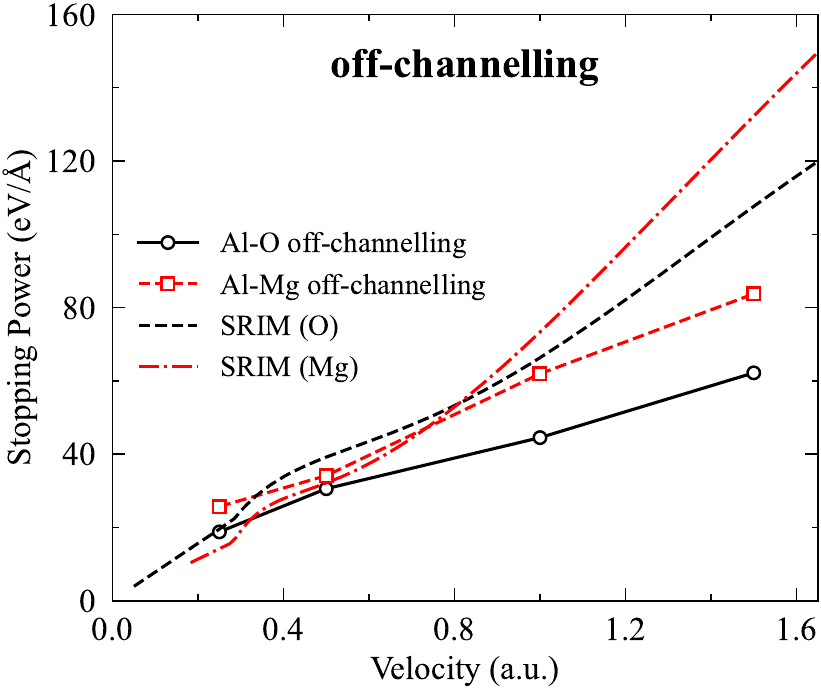}
\caption{$S_e(v)$ curves for O (black) and Mg (red)
along incommensurate averaging trajectories.}
\label{fig:off-channelling}
\end{figure}

  There are trajectories, however, for which the curves never cross,
  with $S_e^{\mathrm{Mg}}>S_e^{\mathrm{O}}$ at all velocities
(the 001-c, 111-c, and incommensurate trajectories among the 
ones tried).
  It happens for trajectories with small impact parameters,
where projectiles pass close to host nuclei 
(see Fig.~\ref{fig:non-crossings}).
  Therefore, an important result is that the $Z_1$ oscillation
as such is trajectory dependent.
  Indeed, the off-channelling trajectory also samples 
close collisions and displays no crossing, as shown
in Fig.~\ref{fig:off-channelling}.
  It is likely that the incommensurate off-channelling trajectory
is not giving an adequate average $S_e$, as the discrepancy 
with SRIM would seem to indicate, but the result is 
consistent with the other small-impact-parameter trajectories. 

  The latter results relate to a low-velocity anomaly that can be
observed in some $S_e(v)$ curves in Fig.~\ref{fig:stopping}:
  The $c$ trajectories both for $\langle 111\rangle$ and 
$\langle 001\rangle$, and both for O and Mg, they all show 
a $S_e$ that does not tend smoothly towards the adiabatic 
limit $S_e(v\rightarrow 0) \rightarrow 0$ on the velocity 
scale presented, not unlike similar anomalies observed 
in insulators or semiconductors \cite{Lim2016,Massillon2025}.
  This will be the focus of a different paper, 
though~\cite{Noor-anomaly}.

  It should also be noticed that the obtained electronic stopping
power values seem to be underestimated when compared to SRIM
and experiment (in Fig.~\ref{fig:stopping}), especially towards
higher velocities.
  Convergence tests with respect to technical parameters and 
basis set were performed, but the discrepancy seems to be more 
fundamental, possibly related to the adiabatic-XC used, or the 
fact that the non-local pseudopotentials were not corrected for 
motion~\cite{Stengel2026}.
  Based on previous experience, however, our expectation is that
such effect is due to the fact that the basis functions for 
core electrons moving with the projectile are missing a velocity 
phase $e^{im\mathbf{v} \cdot \mathbf{r}/\hbar}$.
  If so, it could be easily fixed by resorting to a plane-wave method,
but at a great computational expense. 
  However, for the questions addressed here and velocity regime of 
interest in this study, the present results are perfectly adequate.

\subsection{Projectile Charge}
\label{sec:charge}

\begin{figure}[b!] 
\includegraphics[width=0.35\textwidth, trim=0 -5 0 0,
        clip]
  {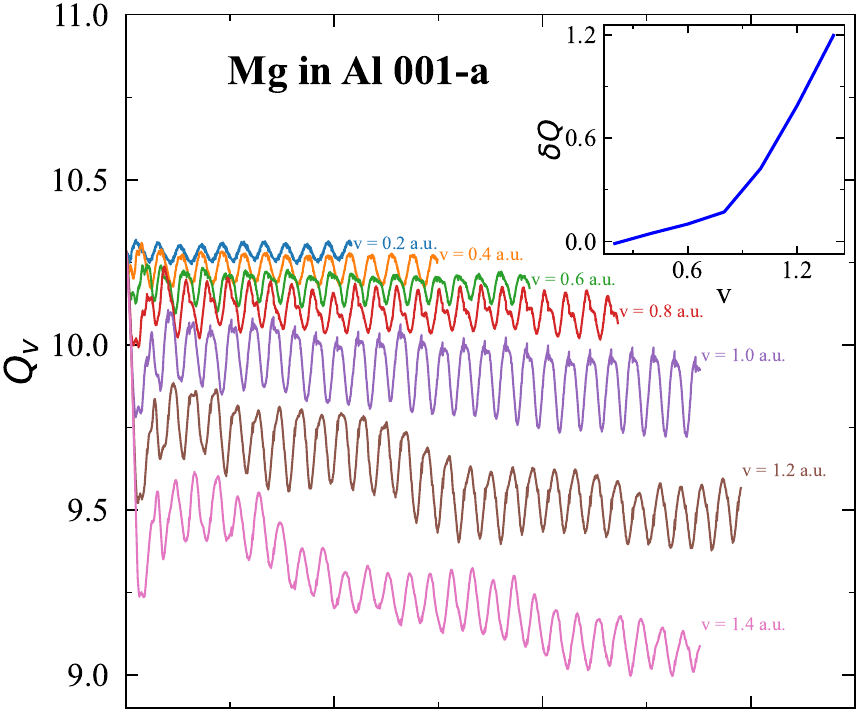}
\vspace{2pt}
\includegraphics[width=0.35\textwidth]
   {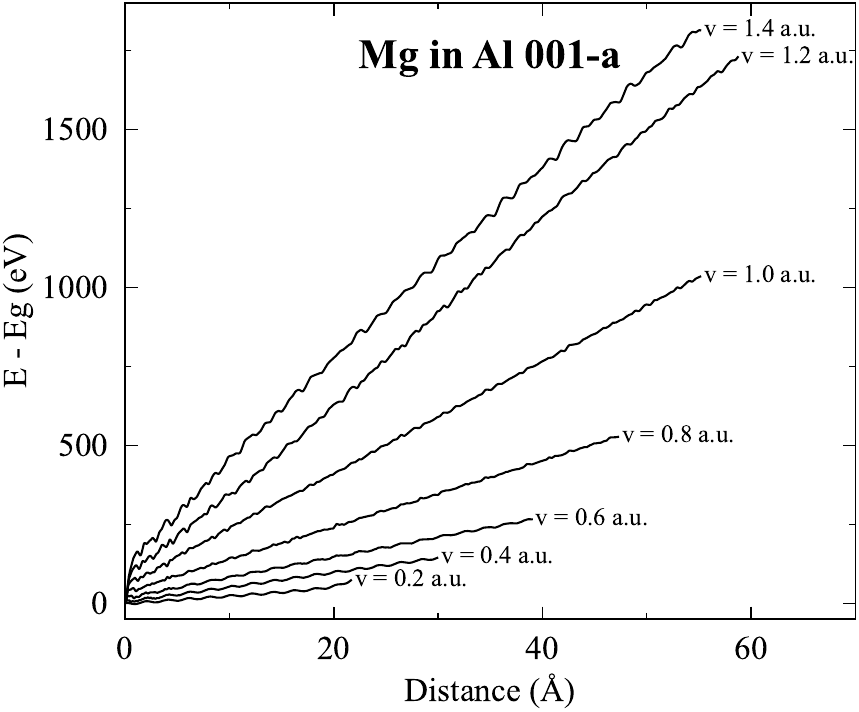}
\caption{Upper panel: Voronoi electron population $Q_v$ for Mg 
in Al along the 001-a trajectory for different values of 
velocity as indicated. 
  Inset: asymptotic average minus its low-velocity adiabatic 
limit, $\delta Q \equiv \langle Q_v^a\rangle-\langle Q_v\rangle$,
versus projectile velocity $v$.
  Lower panel: Excess energy vs displacement for the same system.}
\label{fig:charge-vs-x}
\end{figure} 

  The $Z_1$ oscillations can be seen as an oscillation in the
effective projectile charge, which can be empirically defined for 
any species $s$ by referring its electronic stopping power to the 
proton's (H), $Z_1^* = \sqrt{S_e^s/S_e^{\mathrm{H}}}$.
  Here we want to compare it with actual charges to be obtained 
from the charge density.
  The electronic charge associated to a nucleus is a prototypical 
ill-defined concept (arising from an ill-defined question) with a 
long tradition of utility in chemistry and materials science.
  There are different ways to define such atomic charges or 
populations (see Ref.~\cite{Fonseca2004} for a comparative study).
  In this work we tried two schemes with quite different defining 
rational, the one by Hirshfeld~\cite{Hirshfeld1977}, and the one 
obtained by integrating the charge density within the Voronoi 
polyhedron around each nucleus~\cite{Fonseca2004} (notice that 
we integrate the total density, not the deformation density as
preferred in~\cite{Fonseca2004}).
  They do not give the same numbers, of course, but give very 
consistent results, as shown in Fig.~\ref{fig:charge-comparison}
in the Appendix, which supports the use of either for correlating
their qualitative behaviour when changing species and velocity with 
the corresponding one for $S_e$ and $Z_1^*$.
  We will use the Voronoi electron population $Q_v$ henceforth,
except for an orbital decomposition to assess core bases population,
for which we use Mulliken populations~\cite{Mulliken1955}.

  Figure~\ref{fig:charge-vs-x} shows the evolution of $Q_v$
as a Mg projectile displaces along the centre of the 
$\langle 001\rangle$ channel of the Al target.
  Besides the prominent oscillations related to the Al
(001) plane separation, the curves tend to flatten reaching 
lower values for higher velocities, meaning the projectile
becomes more positively charged.
  The inset shows $\delta Q \equiv \langle Q_v^a\rangle-\langle
Q_v\rangle$ versus velocity, with $\langle \dots \rangle$ 
indicating the corresponding asymptotic averages (averaging away 
their respective oscillations), and $Q_v^a$ standing for the 
population in the adiabatic limit for the considered trajectory.
  $\delta Q$ is therefore a measure of the positive 
charge acquired by the projectile as it moves.

\begin{figure}[t!] 
\includegraphics[width=0.35\textwidth]{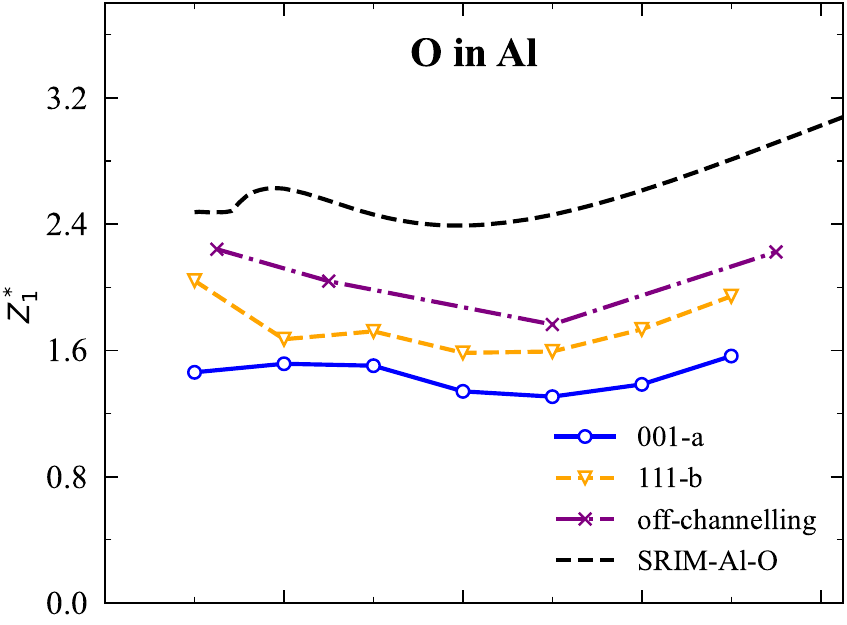}\\ 
\vspace{-3pt}
\includegraphics[width=0.35\textwidth]{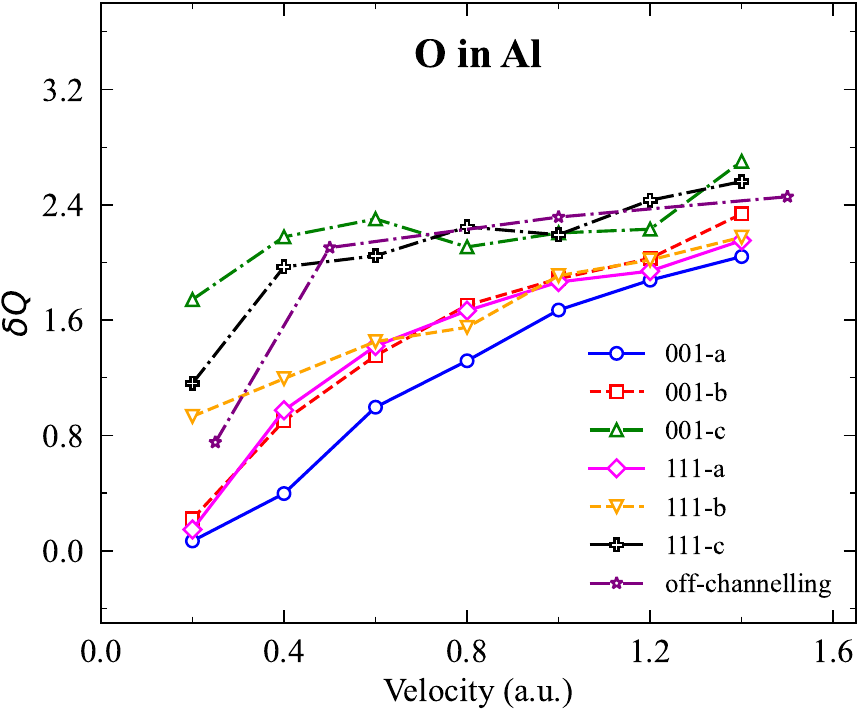} 
\caption{Charge vs velocity for various trajectories of
O in bulk Al as indicated in the key.
Upper panel: Effective charge $Z_1^*$.
Lower panel: Voronoi charge.}
\label{fig:charge-O}
\end{figure} 

\begin{figure}[t!] 
\includegraphics[width=0.35\textwidth]{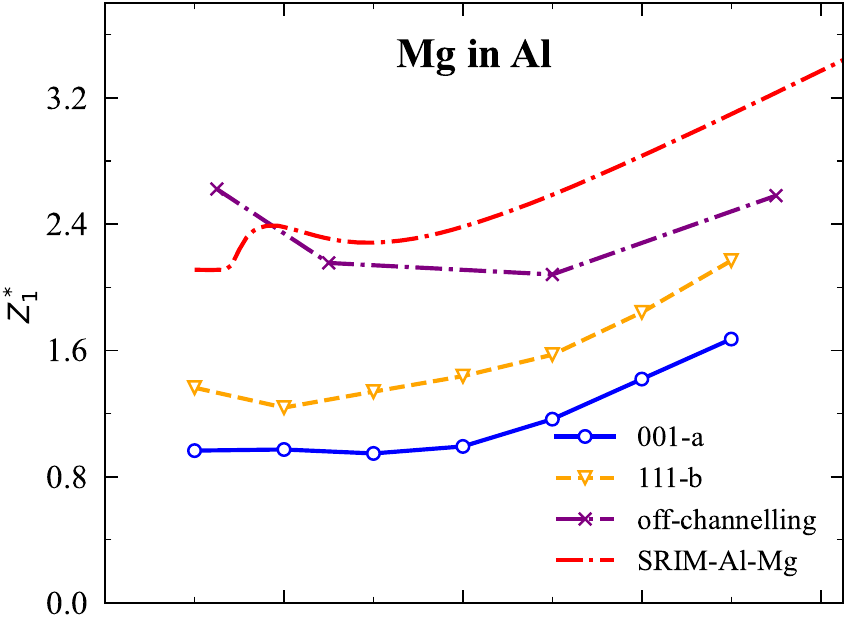}\\ 
\vspace{-3pt}
\includegraphics[width=0.35\textwidth]{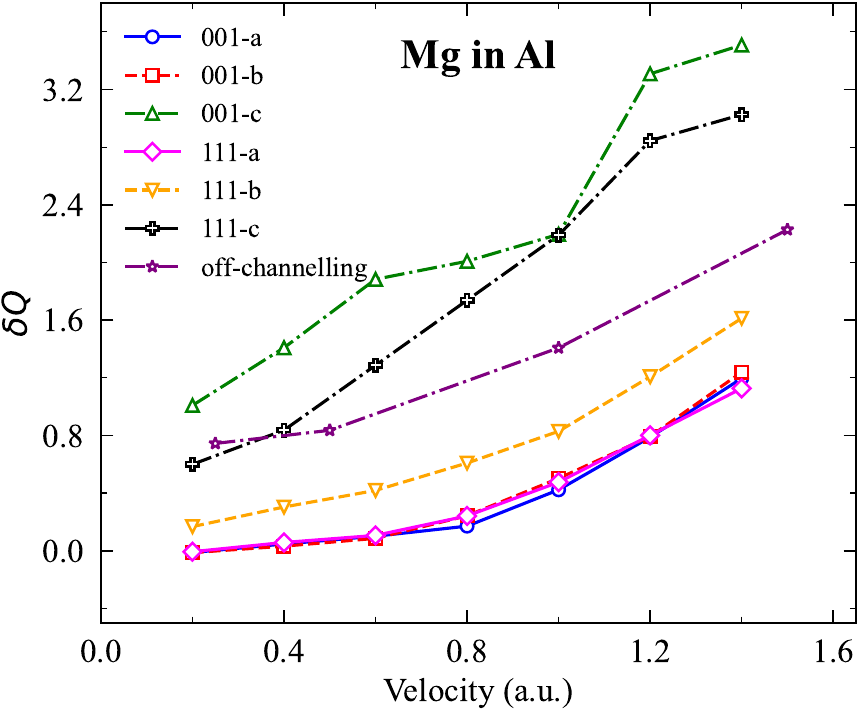} 
\caption{Charge vs velocity for various trajectories of
Mg in bulk Al as indicated in the key.
Upper panel: Effective charge $Z_1^*$.
Lower panel: Voronoi charge.}
\label{fig:charge-Mg}
\end{figure} 

  It is interesting to note the significant distance
travelled before the extra charge $\delta Q$ stabilises.
  It is attributed to the transient behaviour after the
abrupt onset of motion of the projectiles in the 
simulation~\cite{Santervas2025}. 
  At the highest velocity presented (1.4 a.u.) it is
slightly under 60~\AA, which corresponds to 81 a.u. of time 
(2 fs).
  It is quite remarkable when comparing it with the transient
experienced by the energy. 
  The lower panel of Fig.~\ref{fig:charge-vs-x} shows that
the slope of the energy increase stabilises around 
$\sim 6$~\AA\ for the same trajectory, projectile and velocity.
  In previous works~\cite{Halliday2022} a reasonably stationary 
behaviour for the density was reported after the energy 
transient.
  Here, the particular characterisation of the density given 
by $Q_v$ seems to require significantly longer than the 
stabilisation of the energy (the Hirshfeld population offers 
a very similar transient to Voronoi's). 
  It is interesting, however, that for the higher velocities, 
for which $\delta Q$ takes longer to stabilise, there is a 
noise in the $\Delta E(v)$ curves analogous to what found 
for the quantum kick~\cite{Santervas2025}, superimposed 
to the stopping slope and the crystalline oscillations 
better defined at the lower velocities.
  But there is no net increase of stopping power in the 
6 \AA\ to 60 \AA\ regime in which the projectile increases its
positive effective charge.
  The transient behaviour may be worth further exploring in
future work. 

  Figures~\ref{fig:charge-O} and \ref{fig:charge-Mg} compare
the velocity dependence of $Z_1^*$ and $Q_v$ for O and Mg
projectiles, respectively.
  $Z_1^*$ curves were obtained by computing the electronic 
stopping power for protons at the corresponding conditions
and trajectory for each curve.
  The behaviour of $Z_1^*$ is as expected from $S_e$, with
($i$) reasonably flat curves at low velocities, offering 
very similar values for Mg and O, slightly lower for
Mg in the case of SRIM and some trajectories, but not for all;
($ii$) an increase in $Z_1^*$ at velocities $v\gtrsim 0.8$ a.u. 
more pronounced for Mg; ($iii$) a significant trajectory dependence;
and ($iv$) an upturn in the low velocity limit for trajectories
of low impact parameter, related to the mentioned low-$v$ anomaly.

\begin{figure*}[t!] 
\raggedright
\includegraphics[width=0.96\textwidth]{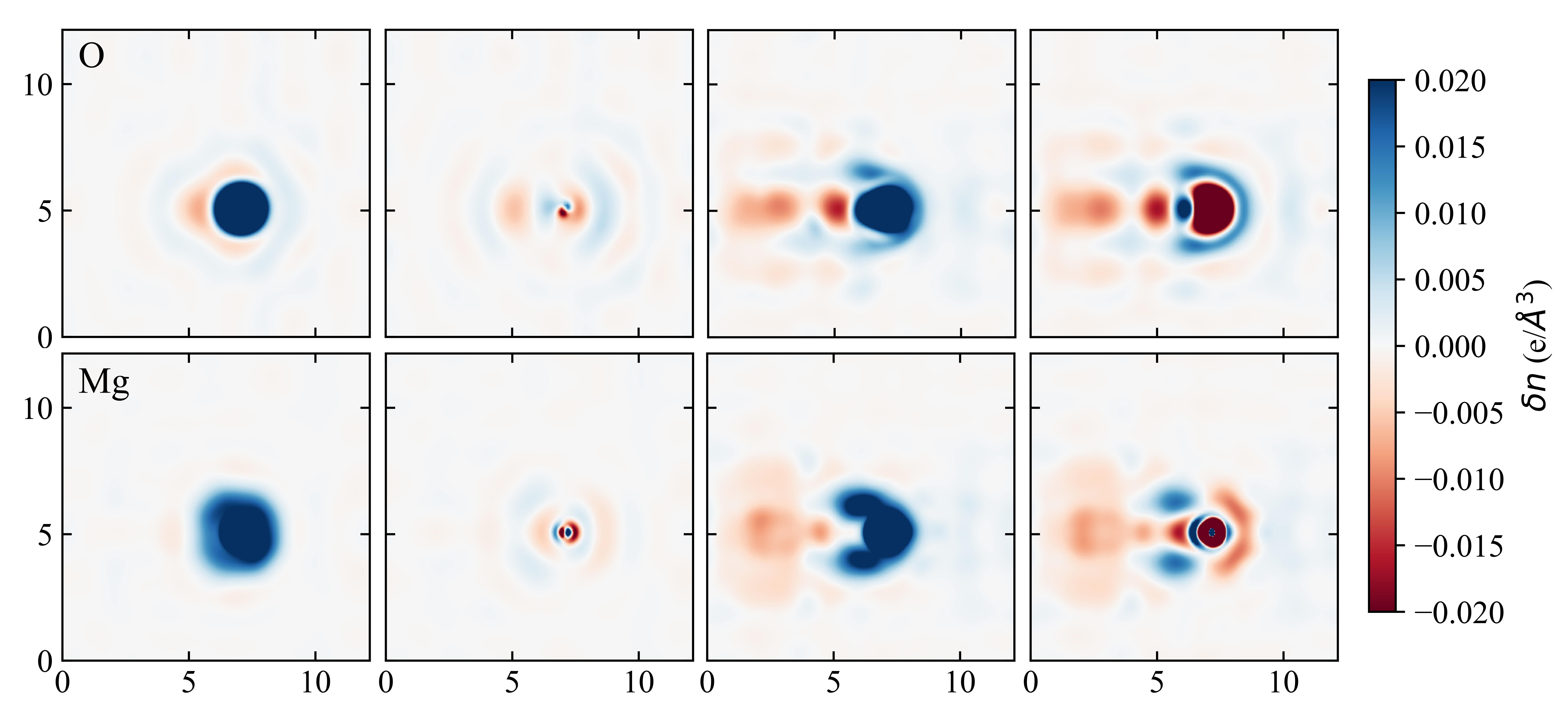} \\
\vspace{-6pt}
\includegraphics[width=0.98\textwidth]{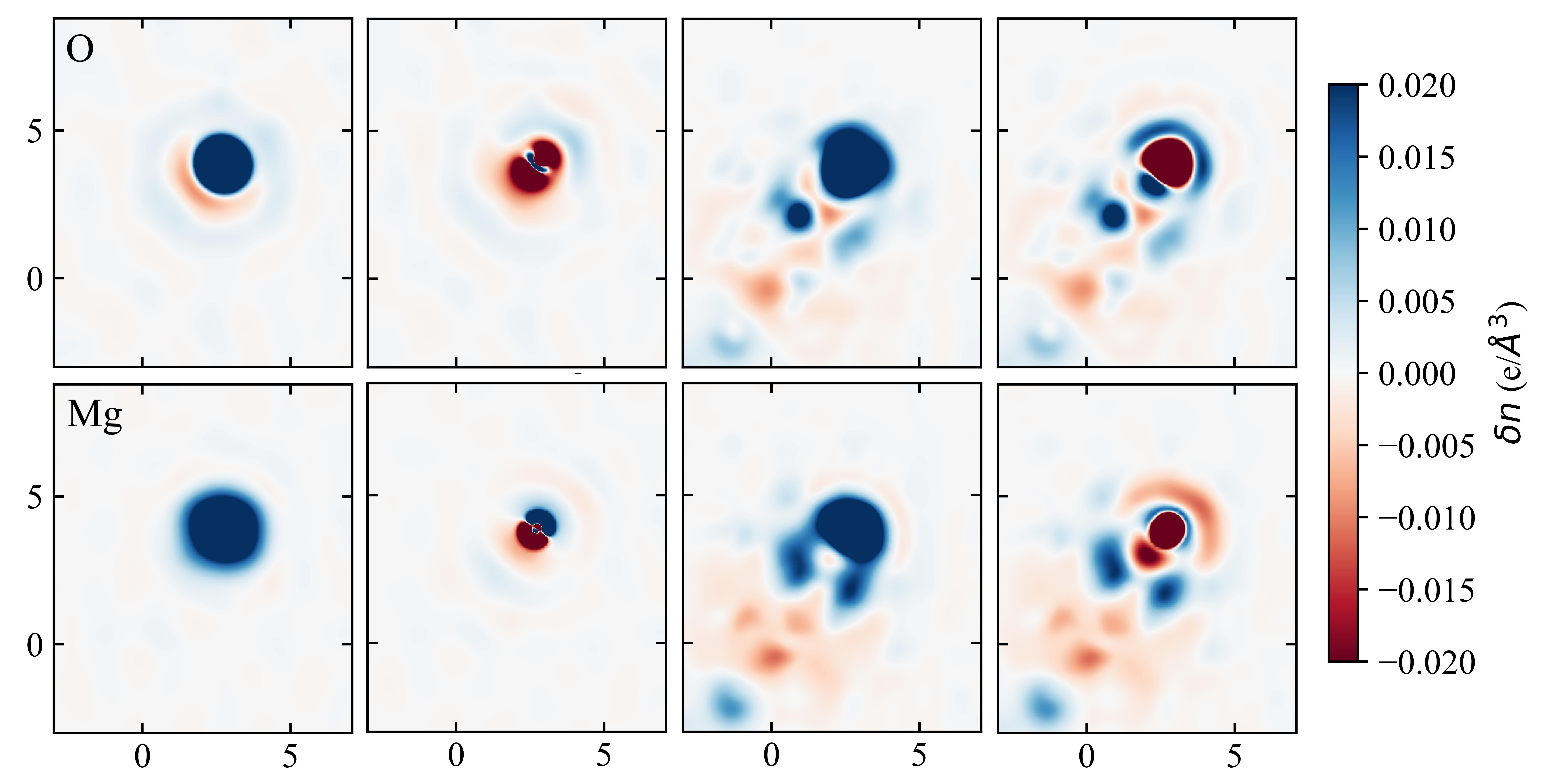} \\
\vspace{-6pt}
\caption{Deformation density maps for O and Mg 
projectiles moving across bulk Al along the 001-a (upper eight panels)
and 111-b (lower eight panels) trajectories, depicted for 
trajectory-containing (010) and $(1\bar{1}0)$ planes, respectively.
  Rows 1 and 3 for O, 2 and 4 for Mg.
  The first two panels in each row correspond to $v=0.2$ a.u.,
the last two for $v=1.2$ a.u.
  First and third panels in a row show $\delta n_b(\mathbf{r},t)$,
second and fourth $\delta n_a(\mathbf{r},t)$.}
\label{fig:dyn-density}
\end{figure*} 

  The $\delta Q$ curves for the corresponding trajectories
shown in the lower panels of Figs.~\ref{fig:charge-O}
and \ref{fig:charge-Mg} are very different, though.
  They tend to zero in the adiabatic limit by definition
of $\delta Q$ (interestingly \cite{Noor-anomaly}, the ones 
that go towards zero more abruptly are the same ones showing 
a low-$v$ anomaly for $S_e$).
  The trajectory-dependent values of the adiabatic 
populations used as reference, $\langle Q_v^a\rangle$, 
are not informative in the context of this paper. 
  For Mg along 001-a (Fig.~\ref{fig:charge-vs-x}), for
instance, $\langle Q_v^a\rangle=10.3$, which just indicates 
that a static ion in a metal is surrounded by electrons 
screening its charge.
  Indeed, it was never expected that an electronic 
population of a static nucleus in a metal would give
anything close to the net charge of the nucleus and
core electrons.
  The proton itself as projectile in Al shows a 
static Voronoi population of $\sim 1.5$.   
  The key point, however, as hinted already in
Fig.~\ref{fig:charge-vs-x}, is that the net charge
grows significantly with velocity, with $\delta Q$ 
increasing between $+1e$ and $+3e$ depending on trajectory,
for Mg at $v=1.3$ a.u., much more than the change 
of $Z_1^*$.
  For O the charging at higher velocities follows a
similar trend, although with a smaller trajectory 
dependence of the net charge at the high end of the 
computed velocities, which is attributed to the fact that
only small impact parameters affect the core electrons
of Mg.
  H charges by $+0.5e$ in the same range.
  
\begin{figure*}[t!] 
\includegraphics[width=0.98\textwidth]{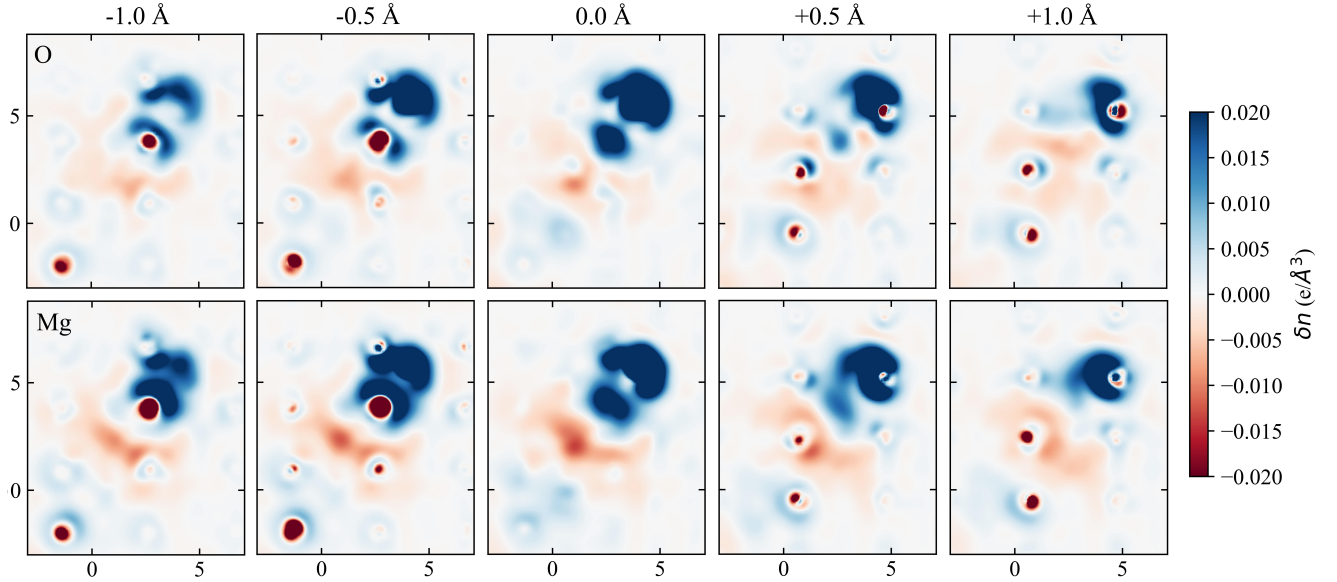} \\
\vspace{-6pt}
\caption{Maps of bulk deformation density $\delta n_b(\mathbf{r})$
for O (upper panels) and Mg (lower panels) projectiles crossing 
through bulk Al along the 111-b trajectory at
$v=1.2$ a.u. and a displacement of 9.44 \AA.
  $\delta n_b(\mathbf{r})$ is shown on $(1\bar{1}0)$ planes displaced 
from the one containing the trajectory by $-1.0$ \AA, $-0.5$ \AA, 0.0 \AA,
0.5 \AA and 1.0 \AA\ as indicated.}
\label{fig:dyn-density2}
\end{figure*} 

  The main conclusion at this stage is that, although 
the projectile does get more charged with velocity,
the actual charge increase coming from the density
and the effective charge defined from stopping power are
quite different magnitudes. 
  The latter is rather a measure of electrons being
involved in the dissipation.

\subsection{Electronic particle density}
\label{sec:density}

  The time-dependent electron density of the evolving 
system $n(\mathbf{r},t)$ offers a much richer information 
than the populations presented above.
  Fig.~\ref{fig:dyn-density} shows heat maps of convenient 
slices of the deformation density for O and Mg projectiles 
propagating through bulk Al.
  The 001-a and 111-b trajectories are presented on 
respective (010) and $(1\bar{1}0)$ planes containing them.
  Two velocity values are shown, 0.2 a.u. and 1.2 a.u., to see 
the differences between the low velocity limit displaying 
the $Z_1$ oscillation and a velocity after the crossover, 
where the oscillation has disappeared.
  Fig.~\ref{fig:dyn-density} shows two deformation densities.
  Firstly, it is defined
with respect to the bulk Al density: $\delta n_b
(\mathbf{r},t) = n(\mathbf{r},t) - n_b(\mathbf{r})$, 
with $n_b(\mathbf{r})$ being the ground-state particle
density of bulk Al without the projectile. 
  Secondly, it is the deformation density with respect to 
the adiabatic density, $\delta n_a (\mathbf{r},t) = 
n(\mathbf{r},t) - n_a(\mathbf{r}, \mathbf{R}(t))$, 
where $n_a(\mathbf{r},\mathbf{R}(t))$ represents the 
ground-state electron density for bulk Al and the projectile 
fixed at the position it would have at the corresponding 
time, $\mathbf{R}(t)$.
  All densities are depicted at the time at which the projectile 
was displaced by 7.08 \AA\ from its starting position.

  The projectile position is the same in all the 001-a maps. 
  It is also the same in all the 111-b maps. 
  Both positions are clearest in the panels of 
Fig.~\ref{fig:dyn-density} for $\delta n_b(\mathbf{r})$ for 
O at $v=0.2$ a.u., sitting close to the centre of the 
saturated positive density circle.
  The mentioned feature shows the excess density due to the
eight extra O electrons travelling with the O ion (nucleus 
and 1$s$ core electrons, replaced by the pseudopotential).
  $\delta n_a(\mathbf{r})$ shows the distortion due to the 
motion, being quite small for $v=0.2$ a.u. for both O and Mg,
especially for the 001-a hyperchannelling case. 
  For 111-b, a partial depletion appears for O, 
consistent with the Voronoi $\delta Q$ appearing in 
Fig.~\ref{fig:charge-O} for that trajectory and velocity.
  The missing charge is mostly accumulated in a positive 
density (blue) around the ion region.

  The comparison with the corresponding low-velocity panels 
for Mg offers relevant insights. 
  $\delta n_b(\mathbf{r})$ shows the Mg electrons around the
ion, with a wider spread than for O as corresponds to the
3$s$ valence orbital of Mg, while still accumulating the 
eight semicore electrons more tightly bound to the ion.
  Interestingly, $\delta n_a(\mathbf{r})$ for Mg is quite
different from that for O, with a smaller redistribution
of the charge along both trajectories, consistent again with 
the low $\delta Q$ appearing in Fig.~\ref{fig:charge-Mg}.
  The 111-b trajectory does show a small but nicely
defined $p$-like distortion on the scale of the Mg 2$p$ 
orbitals indicating a small involvement of the highest 
semicore electrons due to the smaller impact parameter
than in (001)-a.
  Overall the maps reflect the fact that, except for the latter
effect, Mg's semicore electrons are quite rigidly moving with 
the ion at low velocities, consistent with both a good 
screening of the ion change and their barely participating 
in the dissipation, giving a low effective charge $Z_1^*$.
  The fact that it is not so much lower than O's as a naive
valence-electron interpretation would give reflects the fact
that core electrons do have some participation.

  The situation is quite different at the higher velocity.
The deformation is richer for both projectiles, with 
a ripple in front and a wake developing behind, with 
clear differences between both projectiles. 
  Importantly, however, Mg invokes now as much deformation 
$\delta n_a(\mathbf{r})$ as O if not more, meaning that 
Mg's semicore electrons are now well perturbed by the motion, 
consistently with their being engaged in the dissipation.

  The fact that the change in behaviour with increasing velocity
is due to semicore electron participation is in accordance
to the fact that the low-velocity $Z_1$ oscillations follow
the periodic table (and valence electrons) on the one hand, 
and the fact that $S_e$ at the Bragg peak is dominated
by core electrons \cite{Ullah2018}, on the other.
  The slope increase of the $S_e(v)$ curve for Mg in 
Fig.~\ref{fig:SRIM}, is therefore ascribed to the 
incorporation of Mg's $2s$ and $2p$ electrons (mostly the latter).
  It can also be related to a similar upturn observed for light
projectiles in noble metals~\cite{Markin2009,
Zeb2012, Goebl2013} due to a $d$-electron onset of dissipation
beyond the low-velocity limit involving only valence $s$
electrons.
  In this case the upturn is due to semicore electrons
of the target, but the principle is the same.

\begin{figure}[t!] 
\includegraphics[width=0.35\textwidth]{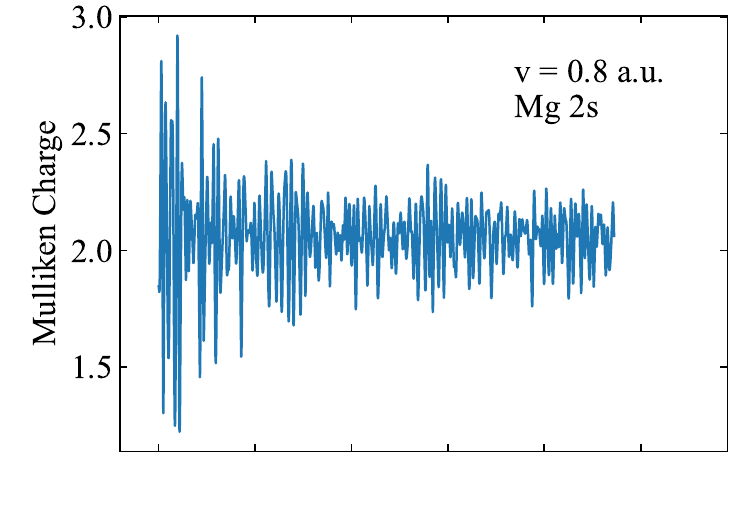} \\
\vspace{-16pt}
\includegraphics[width=0.35\textwidth]{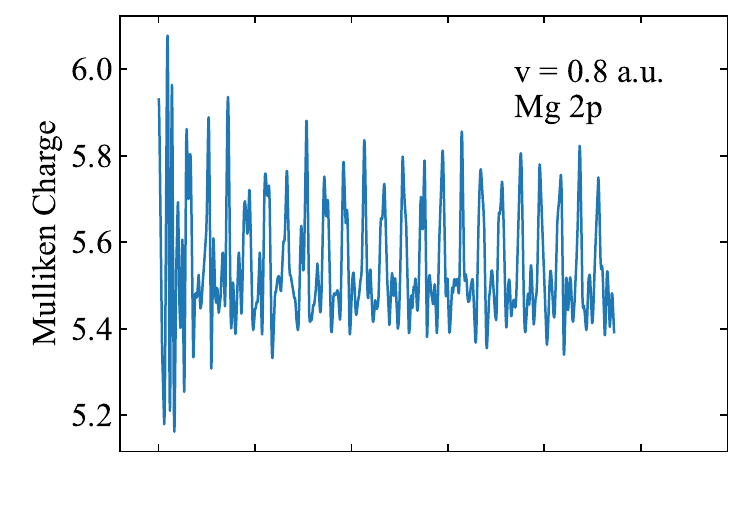} \\
\vspace{-16pt}
\includegraphics[width=0.35\textwidth]{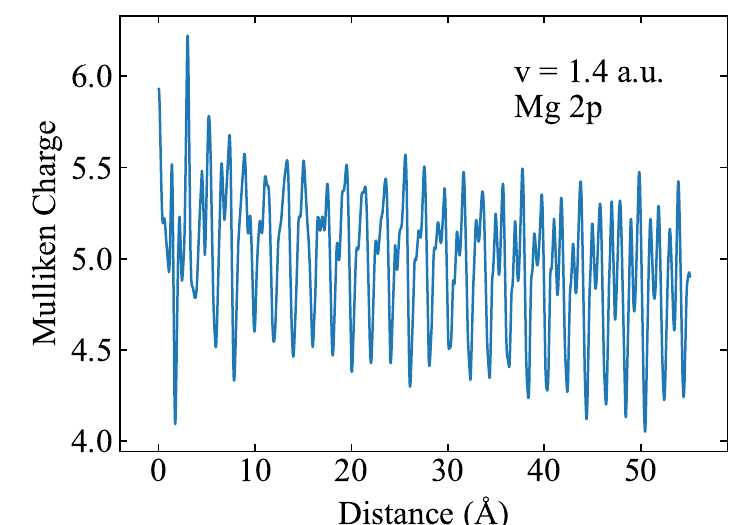} \\
\vspace{-2pt}
\caption{Evolution of population of Mg semicore $2s$ and
$2p$ states for Mg in Al along the 001-a trajectory.}
\label{fig:mulliken}
\end{figure} 

  Following the last point it is important to note
that the participation of the projectile's deeper
electrons in the dissipation correlates with the 
excitation of semicore electrons of the Al target.
  This is appreciated in Fig.~\ref{fig:dyn-density2} where
heat maps for the 111-b trajectory for O and Mg are
shown in $(1\bar{1}0)$ planes parallel to the one 
containing the trajectory at various distances.
  The closest Al atoms to the trajectory sit on the
plane at distance of 0.672 \AA\ from the trajectory,
between the planes indicated by $-1.0$~\AA\ and $-0.5$~\AA\ 
in the figure (the latter being the closest).
  The corresponding maps show two clear depletion (red)
spots (most prominent in the $-0.5$~\AA\ one) corresponding
to the two Al atoms closest to the path.
  The depletion exist for both O and Mg, more marked for
the latter, consistently with Mg semicore states inducing a
more pronounced Al semicore excitation in close collision.
  The planes on the other side of the trajectory path
(at $+0.5$~\AA\ and $+1.0$~\AA) show smaller but noticeable
depletion of Al atoms on the two (111) atomic planes between 
the ones for the closest encounters.

  It is, finally, illustrative to go back to populations
focussing now on core states.
  For this we resort to Mulliken population
analysis~\cite{Mulliken1955}, specifically the populations 
for Mg's $2s$ and $2p$ basis states.  
  Their absolute values are not very reliable in general,
but their variations are good qualitative trend indicators.
  Their time dependence is shown in Fig.~\ref{fig:mulliken} 
and they do show interesting trends.
  The upper panel shows the $2s$ population for Mg along
the (001)-a trajectory at $v=0.8$ a.u., which is close to the
upturn onset for that trajectory (see Fig.~\ref{fig:crossings}).
  The oscillations are due to the sudden start of projectile 
motion in the simulations~\cite{Santervas2025} (quantum kick or 
Migdal effect\cite{Migdal1939}), which tend to thermalise
away.
  The average occupation remains around 2, however.
  
  Compare it with the lower panel of Fig.~\ref{fig:mulliken}
for $2p$ populations at $v=1.4$ a.u., well after the 
crossover.
  The oscillations present a dominant component with a 
frequency corresponding to $\sim 64$ eV, coinciding with
the binding energy of the $2p$ states in (our DFT's) Mg.
  Importantly, however, the amplitude does not decay, 
meaning that the oscillations are fuelled by the dissipation
beyond the quantum kick, and the population itself decays,
implying an excitation process.
  The middle panel shows the populations for $2p$ at $v=0.8$
a.u., which shows the features of the lower panel in an
incipient manner.

\begin{table*}[t!]
\label{tab:basis}
\caption{Parameters defining the basis sets used. 
All radii are given in Bohr.
$r_{c1}$ is the finite-support radius for the first $\zeta$.
Each radial function is generated under a soft confinement
potential that diverges at $r_{c1}$ and is flat below $r_c*$
following the prescription in \cite{Junquera2001}, with 
a soft-confinement energy parameter of $V_0= 50$ Ry.
$r_{c2}$ gives the confinement radii for the second $\zeta$
following the split-norm scheme of \cite{Artacho1999}.}
\begin{ruledtabular}
\begin{tabular}{l|ccc|ccc|ccc|ccc|cc}
& & 2$s$ & & & 2$p$ & & & 3$s$ & & & 3$p$ & & \multicolumn{2}{c}{3$d$} \\
& $r_{c1}$ & $r_{c2}$ & $r_c^*$ & $r_{c1}$ & $r_{c2}$ & $r_c^*$
& $r_{c1}$ & $r_{c2}$ & $r_c^*$ & $r_{c1}$ & $r_{c2}$ & $r_c^*$
& $r_{c1}$ & $r_c^*$ \\
\hline
O  & 4.57 & 2.43 & 4.11 & 7.01 & 3.00 & 6.30 &-&-&-&-&-&-& 7.01 & 6.30 \\
Mg & 2.90 & 1.55 & 6.0 & 3.37 & 1.45 & 6.0 & 9.63 & 6.45 & 7.0 & 9.63 & 
6.45 & 7.0 & - & -  \\
Al & 5.0 & 1.3 & 4.5 & 5.0 & 1.38 & 4.5 & 7.50 & 3.77 & 6.75 & 7.01 & 
4.29 & 6.30 & 8.0 & 7.0  \\
\end{tabular} 
\end{ruledtabular}
\end{table*}


\section{Conclusions}

  The relation between $Z_1$ oscillations (and their disappearance) 
with effective charge and charge density has been explored from
first principles for O and Mg projectiles in bulk
Al metal, as prototypes of maximum and minimum of
the first $Z_1$ oscillation and of a simple metal,
respectively.
  Although the established paradigm for $Z_1$ oscillations
relates to projectiles moving through the homogeneous electron
liquid, we find an important dependence on trajectory
for the oscillation and its disappearance originated by 
the presence of host ions. 

  Hyperchannelling trajectories behave as expected,
with $S_e(v)$ crossovers at velocities not dissimilar
from the one proposed by SRIM.
  But for trajectories sampling closer encounters
the stopping power for Mg is never lower than that 
for O.
  The effective charges $Z_1^*$ obtained from 
first-principles calculations of $S_e$ reflect the
same trends.

  A velocity-dependent effective charging of the projectiles,
as derived from the reduction of electron population around
them, correlates with features in particle density maps
establishing the involvement of semicore
electrons in the dissipative process as the velocity 
increases beyond an upturn in the $S_e(v)$ curves.
  This is directly observed in the change from small
deformation for the density around Mg at low velocities,
corresponding to valence electron involvement alone, 
to sizeable deformation at velocities beyond the upturn.
  The population of the $2s$ and $2p$ functions in Mg
confirm that conclusion.

  Finally, the deformation density induced by the 
projectile is much richer than what could be accounted
for by a simple charge model.
  A description in terms of effective multipoles could
be useful, starting by considering the effect of an
effective dipole~\cite{Alducin2002} in addition to the
monopole.


\begin{acknowledgments}
  We are grateful for discussions with Hongrui Zhang, Runfeng Zhou,
Nuria Santerv\'as and M. Ahsan Zeb.
  We acknowledge United Kingdom's EPSRC funding through
Grant no. EP/V062654/1.
  NUL acknowledges funding from the Higher Education Commission
of Pakistan and the Cambridge Trust, and from the Cambridge 
Philosophical Society and Lucy Cavendish College for travel 
support.
  EA acknowledges funding from the Spanish 
MCIN/AEI/10.13039/501100011033 through grants 
PID2019-107338RB-C61 and PID2022-139776NB-C65, 
as well as a Mar\'{\i}a de Maeztu award to Nanogune, 
Grant CEX2020-001038-M. 
  Calculations were performed in the CSD3 high-performance
computing facility at Cambridge and at Mare Nostrum V in the
Barcelona Supercomputer Centre.
\end{acknowledgments}


\begin{figure}[t!] 
\includegraphics[width=0.35\textwidth]{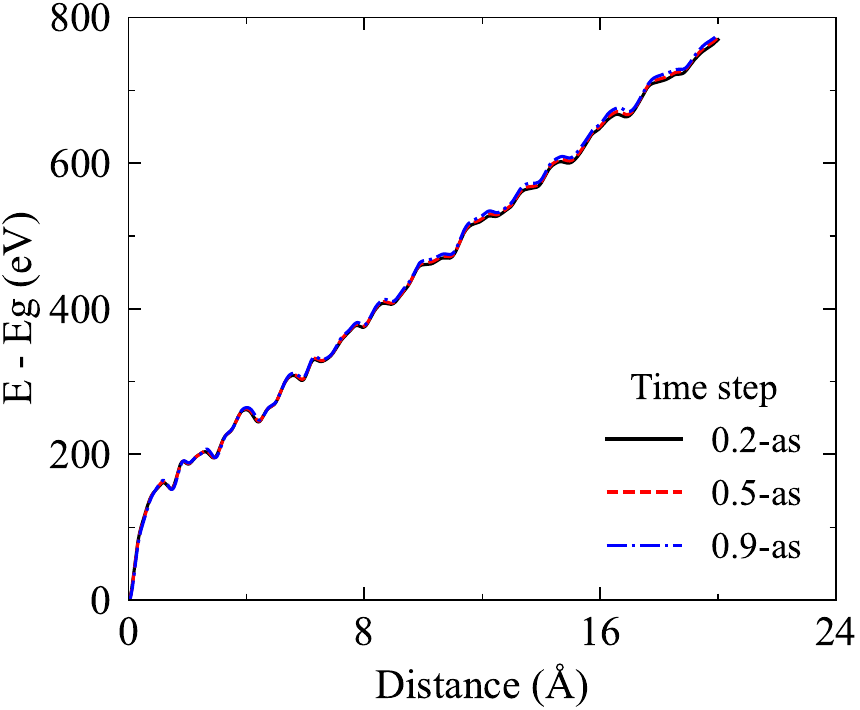} 
\vspace{-6pt}
\caption{Time-step convergence in an excess-energy vs
displacement calculation for Mg moving in bulk Al along
the 001-a trajectory at $v=1.4$ a.u.}
\label{fig:ts-comparison}
\end{figure} 

\begin{figure}[t!] 
\includegraphics[width=0.35\textwidth]
{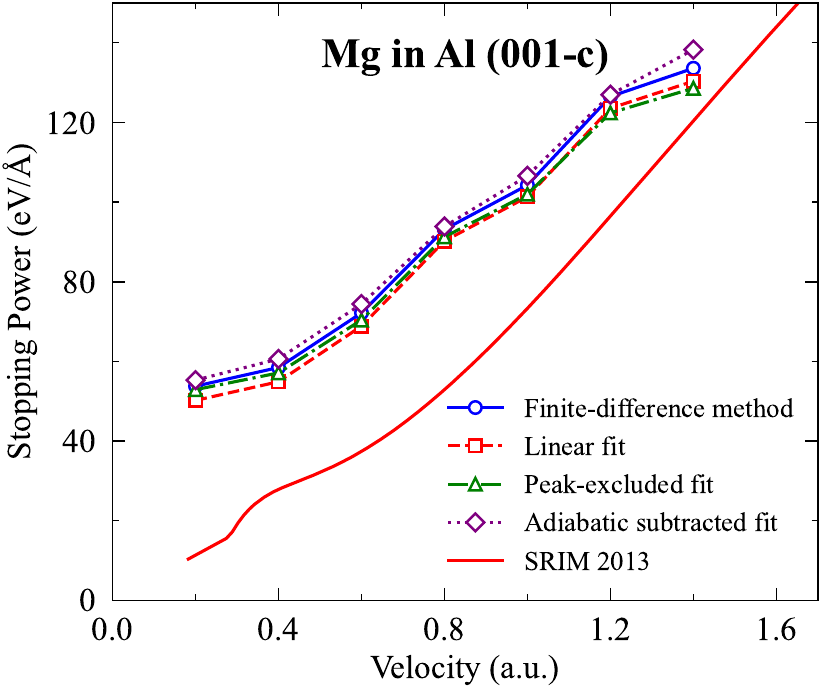} 
\vspace{-6pt}
\caption{Electronic stopping power vs velocity curves 
for Mg in Al along the 001-c trajectory as obtained
from from the excess energy vs distance curves 
at each velocity using different methods of 
obtaining the average slope (after the transient):
Fitting the average value of the derivative of
$S_e$ in a $S_e$ vs displacement $d$ curve obtained
from a finite-difference derivative of $\Delta E(d)$
(blue); a liner fit of the slope directly on 
$\Delta E (d)$ (red); Same but excluding the 
large peaks at close encounters (green); Same as 
red after subtracting the adiabatic energy versus
projectile position along the trajectory (purple).
SRIM-2013 results appear as red continuous line for
reference.}
\label{fig:extractSe}
\end{figure} 

\begin{figure}
\includegraphics[width=0.39\textwidth]
{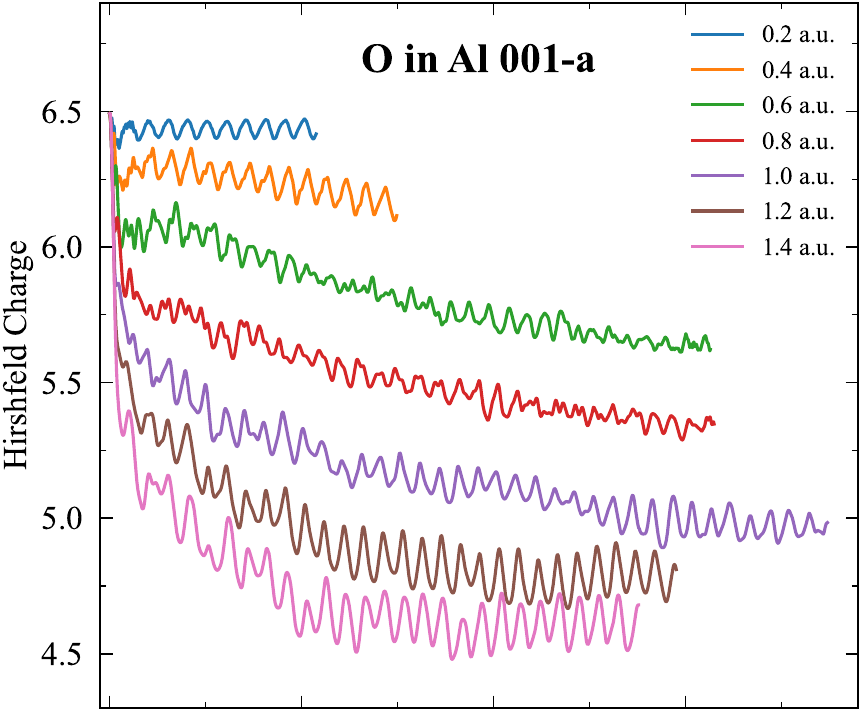} \\
\vspace{1pt}
\includegraphics[width=0.39\textwidth]
{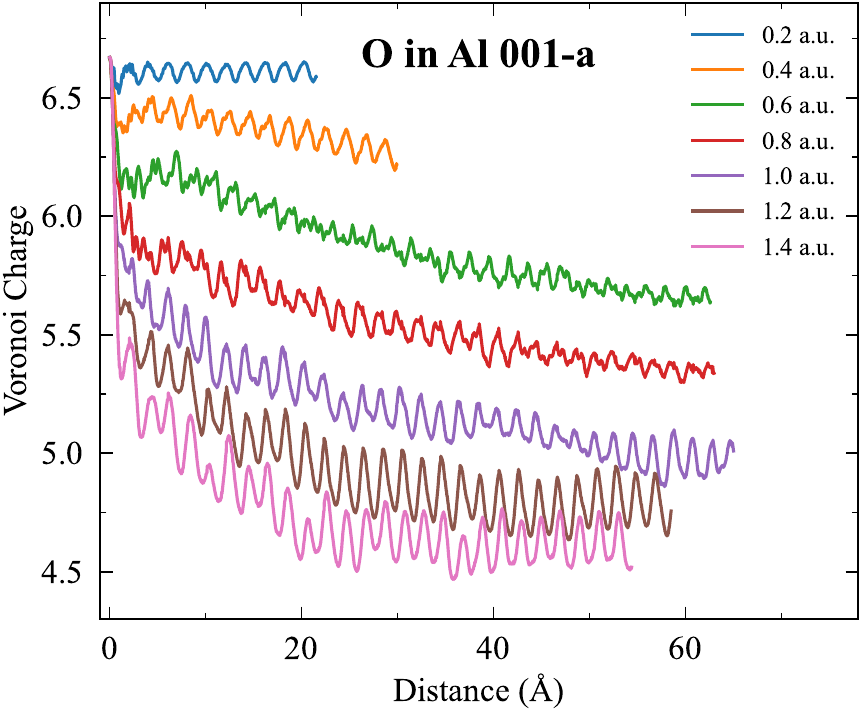} \\
\vspace{-6pt}
\caption{Comparison of Voronoi~\cite{Fonseca2004} 
(upper panel) and Hirshfeld~\cite{Hirshfeld1977} (lower panel)
schemes following the electron population of an O projectile
moving in bulk Al along the 001-a trajectory at 
various velocities as indicated in the key.}
\label{fig:charge-comparison}
\end{figure} 

\appendix*  

\section{Complementary technical data}
\label{appendix}

  The employed basis set is fully specified in Table I. 
  Convergence in time discretisation for the evolution
integrator is presented in Fig.~\ref{fig:ts-comparison}
for the highest projectile velocity used in this
study.
  A time step of 0.9 attoseconds was chosen
for the simulations in this paper.

  There is an ambiguity on how to extract 
$S_e$ from the climb in excess energy with distance.
Fig.~\ref{fig:extractSe} shows the size of the 
error due to this process by comparing the results 
obtained with different methods for the $S_e(v)$
curve for Mg shooting through Al along the 001-c
trajectory, which contains close encounters that
make it more demanding. 
  The curves are compared with the SRIM-2013 curve
for Mg in Al as reference for other errors. 
  Although the differences among methods are noticeable,
they are significantly smaller than difference with
respect to empirical results, and with respect to
trajectory dependence (see Fig.~\ref{fig:stopping}).

  Figure~\ref{fig:charge-comparison} shows the 
qualitative consistency in the assessment of 
projectile population as calculated with the 
very different schemes of integrating the electron
density within the projectile's Voronoi polyhedron
\cite{Fonseca2004} and the Hirshfeld population
scheme~\cite{Hirshfeld1977}.
  Voronoi integration was used in the figures of
the paper.



\end{document}